\documentclass[preprint,secnumarabic,nobalancelastpage,nofootinbibt]{revtex4}

\usepackage{graphics}      
\usepackage{graphicx}      
\usepackage{longtable}     
\usepackage{url}           
\usepackage{bm,color}            
\usepackage{amssymb, amsmath, enumerate, theorem, epsfig,subfigure,setspace}
\usepackage{amsfonts}

\newcommand{\nb}[1]{\textcolor{black}{#1}}

\begin{document}

\title{Towards Arbitrary Control of Lattice Interactions in Nonequilibrium Condensates}
  \author{Kirill P. Kalinin and Natalia G. Berloff}
\email[correspondence address: ]{N.G.Berloff@damtp.cam.ac.uk}
\affiliation{Department of Applied Mathematics and Theoretical Physics, University of Cambridge, Cambridge CB3 0WA, United Kingdom; Skolkovo Institute of Science and Technology, Bolshoy Boulevard 30, bld. 1
Moscow, Russia 121205}


\begin{abstract}{There is a growing interest in investigating new states of matter using out-of-equilibrium lattice spin models in two dimensions. However, a control of pairwise interactions in such systems has been elusive as due to their nonequilibrium nature they maintain nontrivial particle fluxes even at the steady state. Here we suggest how to overcome this problem and formulate a method for engineering reconfigurable networks of nonequilibrium condensates with control of individual pairwise interactions. Representing spin by condensate phase, the effective two spin interactions are created with nonresonant pumping, are directed with dissipative channels, and are further controlled with dissipative gates. The dissipative barriers are used to block unwanted interactions between condensates. Together, spatial anisotropy of dissipation and pump profiles allow an effective control of sign and intensity of the coupling strength between any two neighbouring sites independent of the rest of the spins, which we demonstrate with a two dimensional square lattice of polariton condensates. Experimental realisation of such fully-controllable networks offers great potential for an efficient analogue Hamiltonian optimiser and for reservoir computing.}  

\end{abstract}

\maketitle

{\it Introduction}. In the last decade we have witnessed remarkable achievements in laboratory realisations of exotic states of matter: spin glasses, spin liquids, states of matter with emergent topological order, and gapped or gapless behaviour \citep{SpinLiquid1,SpinLiquid2}. The ultimate goal is to engineer new materials, enter new physical regimes, and develop systematic tools for the controlled preparation and manipulation of various excitations  with applications ranging from fundamental questions in condensed matter physics to analogue Hamiltonian computing. Many spin Hamiltonians have been  implemented and studied using a range of systems: neutral atoms, ions, electrons in semiconductors, polar molecules, superconducting circuits, and nuclear spins among others \citep{QSreview}. Gain-dissipative systems such as photonic and polaritonic lattices have recently emerged as promising platforms for many-body quantum and classical simulations \citep{photonlattices2007,bloch2008}.

When studying  networks with physical systems, one is interested in networks that have  architectures more complex than those of classical random graphs with their Poissonian distributions of connections. For instance, the nature of the spin glass state is well understood for the infinite-range Sherrington-Kirkpatrick model \cite{binder86, mezard87} and  Erd\"os-R\'enyi networks \cite{mezard01}. In a more realistic theory of spin glasses, each spin interacts only with a finite number of neighbours, which brings the notion of dimensionality into the system. Models of this type include the spin glass on a Cayley tree, a Bethe lattice and a disordered random lattice with fixed or fluctuating connectivity. In fact, a finite connectivity structure connects the statistical mechanics of disordered systems with many important optimisation problems, including that of the travelling salesman, graph partitioning, and K-satisfiability problems. The efficient optimization of such computationally hard combinatorial and continuous  problems can be realised by finding the ground state configuration of spin models \cite{lucas}. This can have a profound impact in many areas, e.g. prediction of new chemical materials and various machine learning applications. 

A significant interest is therefore attached to building easily tunable physical systems capable of emulating lattice spin models. In this respect, a variety of physical platforms have been explored including superconducting circuits \cite{supercond}, trapped ions \cite{trapped}, CMOS (complementary metal-oxide-semiconductor) devices \cite{CMOS}, and Coherent Ising Machines \cite{yamamoto14, yamamoto16a, yamamoto16b,takeda18}. These artificial spin networks are constructed and activated in such a way  that they will naturally evolve to the Ising ground state. Compared to  classical optimisation methods that often result in finding a  local energy minima, these novel computing schemes promise a significant speed-up of calculation time for finding the global minimum with respect to classical computing.

We are particularly interested in the new subclass of simulators  based on networks of lasers \cite{coupledlaser} or nonequilibrium condensates such as polariton \cite{NatashaNatMat2017} or photon \cite{KlaersNatPhotonics2017} condensates. These are gain driven systems with dissipation: when gain exceeds the threshold and overcomes the losses  a phase transition to a coherent state characterized by a wavefunction $\psi({\bf r},t)$ occurs. To create a network of condensates -- a condensate lattice -- the gain should occur  at various localised spatial positions within the decoherence length of neighbouring condensates. For example, a common way to experimentally form a polariton lattice is to use a spatial light modulator, which creates polariton condensates at the vertices of any prescribed graph \citep{NatashaNatMat2017,wertz,manni,pendulum,baumbergGeometrical}. Initially each gain centre is seeded with a random phase and is evolved independently of other centres.  As the occupation of each gain centre grows with an  increased pumping intensity, different centres  start interacting while exchanging particles. The dissipative nonlinearity saturates the gain and the system may settle to  a steady state with a particular distribution of phase differences between the network elements. We shall refer to the phase of each gain center, $\theta_i$, as the ``spin" of the $i-$th element in the network.
 Depending on the pumping mechanism (ex. non-resonant or resonant) the phases of the complex amplitudes may take  continuous or discrete values in $[0,2 \pi)$  and  the system may simulate a variety of Hamiltonians. It was shown that non-resonant pumping in coupled lasers \cite{coupledlaser}  or nonequilibrium condensates \cite{NatashaNatMat2017} leads to minimisation of the XY Hamiltonian $H_{XY}=-\sum_{i,j} J_{ij} \cos (\theta_i-\theta_j)$, whereas a combination of resonant and nonresonant pumpings is theoretically predicted to minimise the Ising or $n-$state planar Potts Hamiltonians \cite{KalininPRL18}.  For a general form of the matrix of couplings $J_{ij}$  finding the ground state of such  Hamiltonian  is known to be NP-hard  \cite{ZHANG06}. 
 
In addition to Hamiltonian optimisation, various nonlinear dynamical systems, including 
electronic \citep{electronic_RC1,electronic_RC2,electronic_RC3}, 
photonic \citep{Photonic_RC1,Photonic_RC2}, 
spintronic \citep{Spintronic_RC1,Spintronic_RC2,Spintronic_RC3}, 
mechanical \citep{Mechanical_RC1}, 
and biological \citep{Biological_RC1} systems, 
have been recently employed as potential reservoirs for reservoir computing (RC) (see \citep{Review_physicalRC2018} and references therein). RC methods, originally referred to as echo state networks \citep{ESN_2002} or liquid state machines \citep{LSM_2002}, constitute a computational framework for temporal data processing. These methods have been successfully applied to many practical problems involving real data with most of the studies focused on machine learning applications. The role of the reservoir in RC is to nonlinearly map sequential inputs into a higher-dimensional space so that the features can then be extracted from its output with a simple learning algorithm. Therefore, such reservoirs become attractive for an experimental implementation in many physical systems with a motivation of realising fast information processing devices with low learning cost. Networks of nonequilibrium condensates or lasers can serve as interacting nonlinear elements for an efficient network-type reservoir computing system. In particular, many requirements for a physical RC system can be fulfilled by polariton condensates, and indeed the first proposals of such an implementation have appeared \cite{liewRC}. 
Polariton condensates are scalable to a large number of lattice sites \citep{NatashaNatMat2017}, while high dimensionality is necessary for mapping input data into a high-dimensional space in RC. The polariton network is a strongly nonlinear system \cite{bec}  which rises from the excitonic part of polaritons. This  is necessary for the reservoir to operate as a nonlinear mapping. The presence of short-term memory in polariton systems is supported by many experimental observations of their bistability behaviour \citep{Pickup2018} and such memory is necessary to ensure that the reservoir state is dependent on recent-past inputs but independent of distant-past inputs. Thanks to the polariton's photonic component, the rich physical properties of waves such as interference and synchronization make polariton condensates similar to coupled oscillators \citep{OurLattices2019},  and together with a possible on-chip implementation at room temperature with organic materials, they become a compelling candidate for RC.

In this paper, we study a model of a fully controllable polaritonic network of a fixed geometry by spatially varying dissipation. In this network, the desired interactions between any nodes can be supported by creating channels of low dissipation and further controlled individually by dissipative gates while undesired interactions can be eliminated by high-dissipative barriers. We argue that spatially varying dissipation has an advantage over other methods that involve the changes in the polariton densities at the condensates sites  (e.g. external potential barriers, exciton reservoir injections,  etc.)  and, therefore, change couplings  with other sites in the system. This difficulty is connected to the nonequilibrium nature of condensates that maintain nontrivial particle fluxes even at the steady state. Introducing barriers between the condensates redirects the fluxes and, therefore, affects the densities of the network elements. Dissipative barrier, on the other hand, removes the particles between the neighbouring sites without any profound effect on the site  densities, and, therefore, changes interactions only locally. 

The fixed barriers and dynamically reconfigurable gates can be experimentally imprinted with a technique of proton implanting \citep{MFraser2} or  excited states absorption \cite{RoadMap} as we discuss later in the paper. By taking a square lattice of polariton condensates, we show that full control over the polaritonic network can be reached. The coupling strength of any two adjacent sites can be changed on demand by varying the dissipative gate within a physically reasonable range. We also show that such an approach is easily scalable and for 500 condensates with about 100 barriers a meaningful ground state can be reached, e.g. ``Sk" letters in our case. We justify how such dissipative channels, gates, and barriers, can be experimentally implemented for a few physical platforms, including polariton and coupled laser systems. Such a polaritonic network can be further used for the purpose of RC or analogue Hamiltonian optimisation.

{\it Configuring dissipation for periodic density modulations}. Systems with strong light-matter interactions are potential platforms for the control of spin networks with spatially dependent dissipation profiles. In this paper we consider polaritonic networks made of exciton-polariton condensates and use the mean-field approach to describe polariton condensates, i.e. the generalised complex Ginzburg-Landau equation (cGLE) coupled to the reservoir dynamics \cite{goveq, carusotto, reviewCarusotto}: 
\begin{eqnarray}
	i \hbar  \frac{\partial \psi}{\partial t} &=& - \frac{\hbar^2}{2 m}(1 - i \hat{\eta} n_R)  \nabla^2\psi + U_0 |\psi|^2 \psi+
g_R n_R \psi 	  +\frac{i \hbar}{2} [R_R n_R - \gamma_C] \psi, \\
	\frac{\partial n_R}{\partial t} &=&  - \left(\gamma_R+ R_R|\psi|^2 \right) n_R + P({\bf r},t),
\end{eqnarray}
where $\psi({\bf r},t)$  is the wavefunction of the condensed system that is coupled to the density of the exciton reservoir $n_R({\bf r},t)$ via nonlinear exciton-polariton  interaction strength $g_R$, scattering rate $R_R$, and diffusion rate $\hat{\eta}$ \cite{wouters12,KBuniversal}. The polariton-polariton interaction strength is denoted as $U_0$ and the polariton effective mass as $m$. The polariton and exciton losses are described by  $\gamma_C$ and $\gamma_R$, respectively. The incoherent pump source is described by the pumping intensity $P({\bf r},t)$. We note that although the process of Bose-Einstein condensation of polariton condensates includes quantum effects, once the condensation happened it can be accurately described by the mean-field equations as was shown in numerous experimental studies \cite{pendulum,meanfield2,baumbergGeometrical,meanfield4,meanfield5,meanfield6,meanfield7,meanfield8,meanfield9}.  We nondimensionalise these equations  by $\psi \rightarrow \sqrt{ \hbar R_R / 2 U_0 l_0^2} \psi$, $t \rightarrow 2 l_0^2 t / R_R$, $ {\bf r} \rightarrow \sqrt{\hbar l_0^2 / (m R_R)} {\bf r}$, $n_R \rightarrow n_R / l_0^2$, $P \rightarrow R_R P / 2 l_0^2 $ and introduce the dimensionless parameters  $g= 2 g_R / R_R$, $b_0= 2 \gamma_R l_0^2 / R_R$, $b_1=  \hbar R_R / U_0$,  $\eta= \hat{\eta} / l_0^2$, $\gamma =  \gamma_C l_0^2 / R_R$, where $l_0 = 1 \mu m$. The resulting model yields
 \begin{eqnarray}	
	i  \frac{\partial \psi}{\partial t} &=& - (1 - i \eta n_R)  \nabla^2\psi +  |\psi|^2 \psi+
g n_R \psi 	  + i(n_R - \gamma) \psi, \label{e1}\\
	  \frac{\partial n_R}{\partial t} &=&  - \left(b_0+ b_1|\psi|^2 \right) n_R + P({\bf r},t).
	\label{e2}
\end{eqnarray}

To understand how the presence of spatially varying dissipation affects the landscape and the distribution of the polariton particles we consider the steady state solutions of Eqs.~(\ref{e1}-\ref{e2}) characterized by the chemical potential $\mu$, such that $\psi=\Psi({\bf r}) \exp[i \mu t]$, $\Psi({\bf r})=\sqrt{\rho({\bf r})} \exp[i S({\bf r})]$ and neglect the energy relaxation since $\eta \ll 1$. Equations (\ref{e1})-(\ref{e2}) under these assumptions become
\begin{eqnarray}
\nabla\cdot(\rho u) &=& (n_R - \gamma({\bf r}) )\rho, \label{m1}\\
\mu &=& \rho + g n_R+ u^2({\bf r}) - \nabla^2 \sqrt{\rho}/\sqrt{\rho}, \label{m2}\\
n_R&=&\frac{P({\bf r},t)}{b_0+b_1 \rho}, \label{m3}
\end{eqnarray}
where $u({\bf r})=\nabla S({\bf r}).$ If both the pumping intensity and the dissipation  are  constant both in space and time, there are no velocity fluxes across the sample $u=0$ and both the polariton and exciton reservoir densities take on the constant values $n_R=\gamma$, $\rho=P/b_1\gamma - b_0/b_1$, $\mu=\rho+g\gamma$. Non-uniform pumping profile or non-uniform dissipation lead to nontrivial velocity fluxes and non-uniform densities. We illustrate this by solving Eqs. (\ref{m1})-(\ref{m2}) in the Thomas-Fermi (TF) approximation in 1D.   We neglect the quantum pressure term (the last term on the right-hand side of Eq. (\ref{m2})) and the detuning due to the repulsive interactions between polaritons and hot excitons (the solution method is straightforward to generalize to nontrivial $g$, but leads to more complicated analytical expressions) and look for a self-consistent solution. We shall assume that the velocity profile is periodic with the period  $\beta [\mu m]$: $u(x)=A \sin(2\pi x/\beta)$, therefore, $\rho=1+A^2/2-A^2 \sin^2(2\pi x/\beta),$ where the chemical potential $\mu$ is chosen to normalize $\rho$: $\int_0^\beta \rho \,dx =\beta$ and $A$ is a free parameter. We integrate the quantum pressure term over the period of density oscillations and conclude that the TF condition is met if 
$
A\ll \sqrt{2}(\beta^2 \left(\beta^2+8 \pi ^2\right)/\left(\beta^2+4 \pi ^2\right)^2)^{1/4}.
$
Integrating Eq.~(\ref{m1}) for constant pumping intensity fixes the spatial form of the dissipation
\begin{equation}
\gamma(x)=\frac{2 P}{A^2 b_1 \cos \left(\frac{4 \pi  x}{\beta}\right)+2 b_1+2 b_0}-\frac{2 \pi  A \cos \left(\frac{2 \pi
    x}{\beta}\right) \left(3 A^2 \cos \left(\frac{4 \pi  x}{\beta}\right)-2 A^2+2\right)}{\beta \left(A^2 \cos \left(\frac{4
   \pi  x}{\beta}\right)+2\right)}.
   \end{equation}
    This is a doubly-periodic dissipative structure that creates  periodic density and velocity modulations even in the presence of  a uniform pumping profile. Thus, spatially varying dissipation effectively creates an excitonic landscape of hills and valleys for polariton flows. 

\begin{figure}[t!]
\centering
\includegraphics[width=8.6cm]{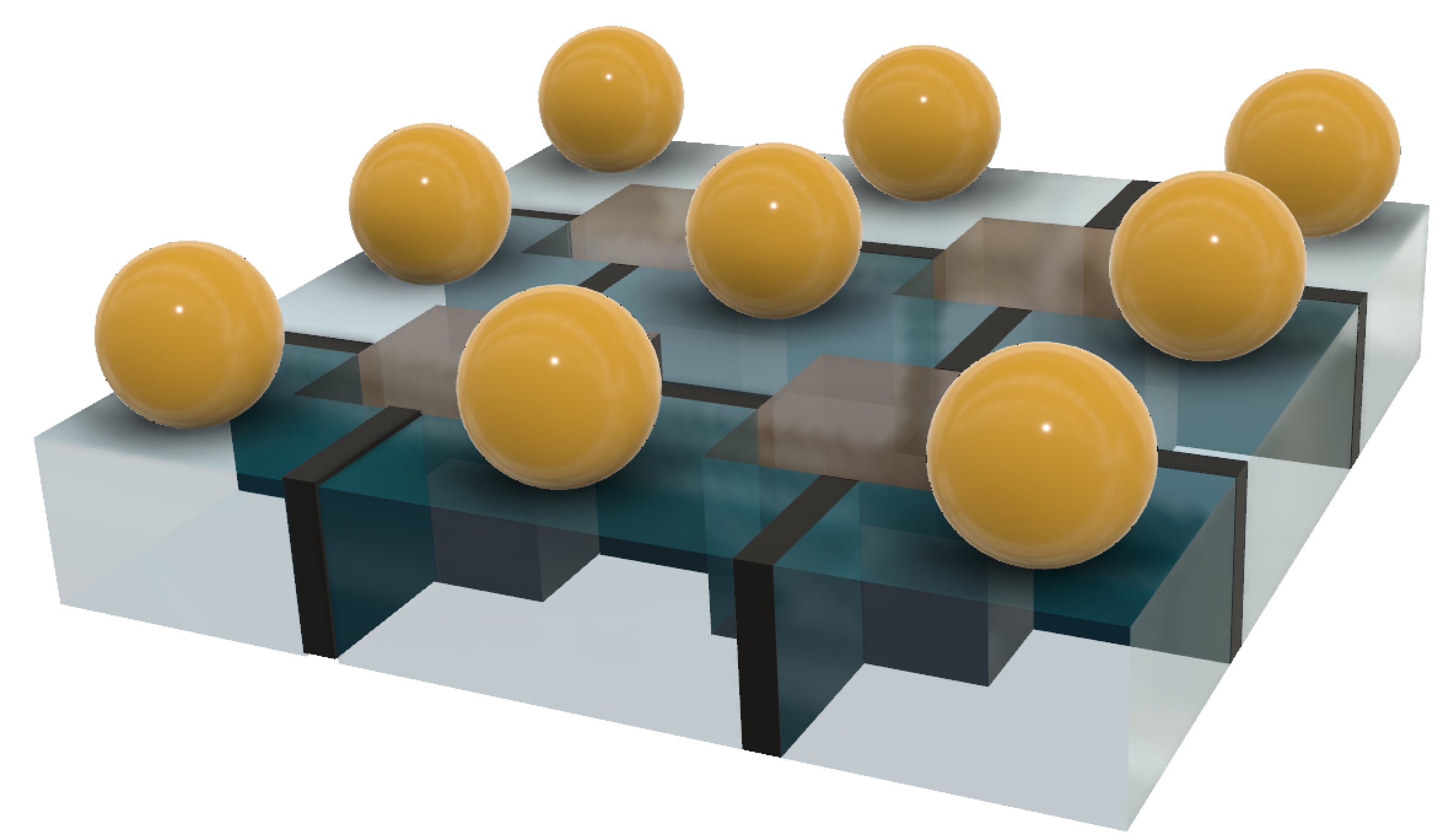}     	
\caption{Schematics of the spatially dependent dissipation profile with nine condensates (yellow balls) arranged in a 2D lattice in a semiconductor microcavity. The condensates interact via channels of low dissipation (light gray). Dissipative barriers (brown areas) show an increased dissipation that prevents the coupling between the condensates across the diagonals. Dissipative gates (dark blue areas) show the areas where the dissipation is increased or decreased to control the couplings between neighbouring condensates.}
 \label{Figure1}
\end{figure}

{\it Control of lattice interactions in 2D lattice}. The analysis of the relationship between density modulations and spatially varying dissipation  suggests to use a varying dissipation profile across the sample to establish pairwise interactions that can be independently tuned  between any two condensates. For instance, this can be achieved by creating a spatially dependent dissipation profile $\gamma({\bf r})$, as illustrated in Fig.~\ref{Figure1}. With this scheme, a two-dimensional square grid is formed. It consists of narrow rectangular strips, i.e. channels, characterised by a constant low  dissipation rate $\gamma=\gamma_{\rm channel}$. Such small dissipation supports flows of polariton quasi-particles through these channels. Outside of the channels, the dissipation is strongly enhanced and equal to $\gamma=\gamma_{\rm barrier}\gg 1$, thereby forming dissipative barriers. The condensates are pumped at the grid vertices (schematically shown as yellow spheres in Fig.~\ref{Figure1}) and the high-dissipative barriers (brown areas) block the outflow of polaritons across the diagonals.  

%
The pairwise interactions between the neighbouring condensates are further controlled by another dissipative layer across the channel: dissipative gates (dark blue narrow blocks in the scheme) given by $\gamma=\gamma_{\rm channel}+\gamma_{\rm gate}({\bf r})$. The amplitude of dissipative gates should be available for a dynamical adjustment and be large enough to change the sign of interactions from ferro- to antiferromagnetic. In contrast, the channel-barrier structure forms the stationary dissipative profile that can be conclusively imprinted in the sample. 
For the simulations below, the physically meaningful polariton lifetimes of $5 ps$, $20-200 ps$, and $200 ps$ (or $2ps$/$13ps$/$100ps$ for simulations in Fig.\ref{Figure4}) are used for barriers, gates, and channels, respectively (with an assumption of the exciton lifetime of $2ns$).

%
\begin{figure}[h!]
\centering
\includegraphics[width=8.6cm]{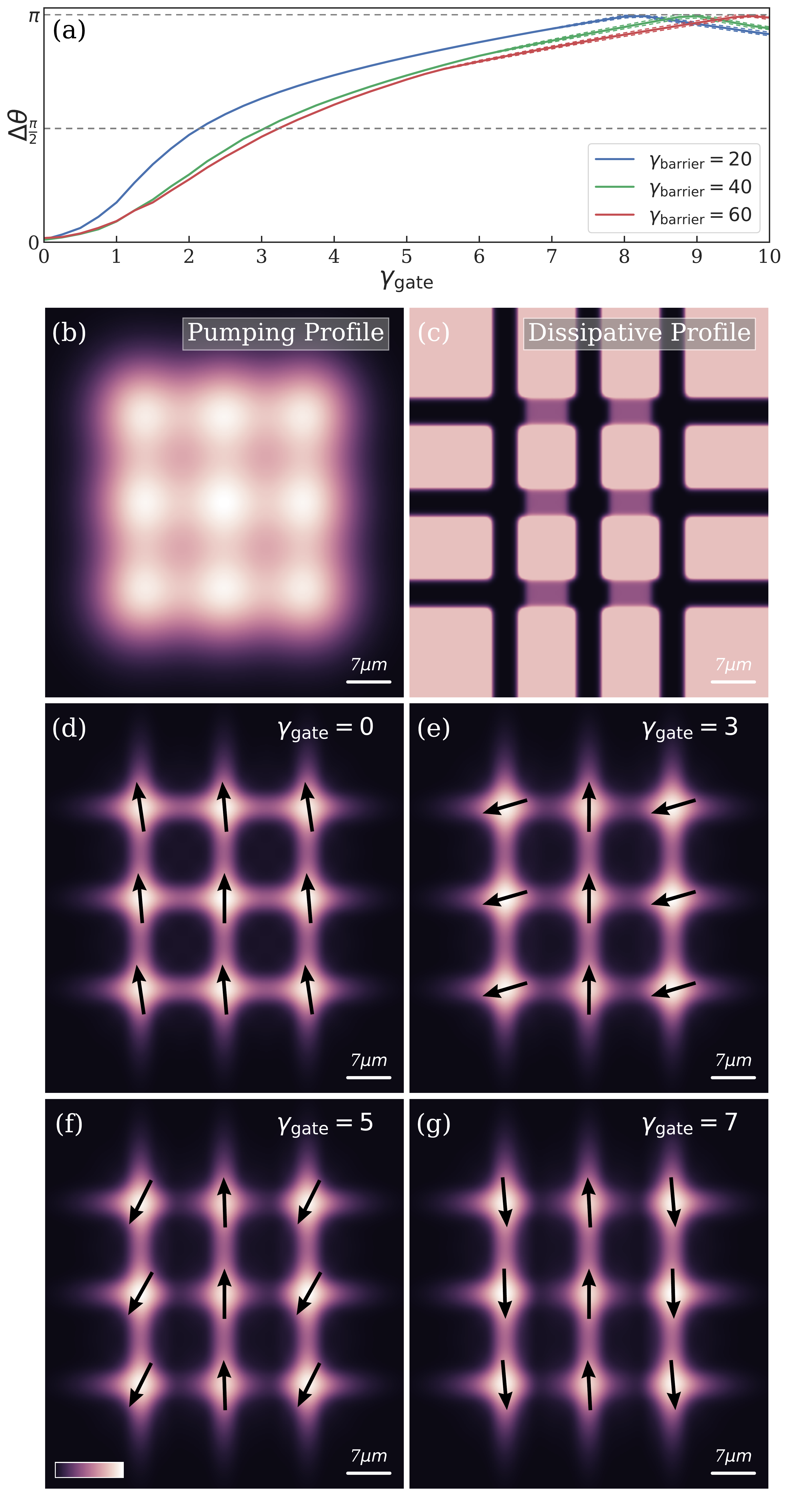}     	
\caption{(a) The transitions from the ferromagnetic state with $0$ phase difference to the antiferromagnetic state with $\pi$ phase difference between two polariton condensates by varying the amplitude of the dissipative gate are shown for three dissipative barriers (solid lines). The condensates remain synchronised though the small phase fluctuations appear for higher dissipative gates. The amplitude of such fluctuations is contained within the dashed lines. (b) The pumping profile for a $3 \times 3$ square lattice of polariton condensates. (c) The structure of a dissipation profile $\gamma ({\bf r})$ consisting of dissipative channels (black), dissipative barriers (pink), and dissipative gates (purple) being horizontally placed between vertical stripes of condensates. (d-g) The phase differences of all condensates are shown with black arrows with respect to the central condensate, the background is the normalised polariton density $|\psi({\bf r})|^2$. (d) In the absence of dissipative gates, the initial state is prepared to be ferromagnetic with all spins aligned in the same direction. (e-g) With a dissipative gate of $\gamma_{\rm gate} = \{3,5,7\}$, the coupling strengths between vertical stripes of condensates are continiously changed from ferromagnetic to antiferromagnetic coupling. Simulation parameters for (\ref{e1},\ref{e2}) are $\eta = 0.01$, $g = 0$ (a small nonzero value of $g \approx 0.1$ will make the spins less aligned), $b_0 = 0.2$, $b_1 = 20$, $ P = \sum_i P_0 \exp(-\alpha ({\bf r - r_i})^2)$, $P_0 = 100$, $\alpha = 0.4$, $\gamma_{\rm channel} = 1$, $\gamma_{\rm barrier} = 40$, lattice constant is $d = 2.8$, the width and length of channels are $1$ and $1.4$. }
 \label{Figure2}
\end{figure}
%

Spatially varied dissipation profiles can lead to the realisation of independent control of individual coupling strengths. This in turn will open unprecedented opportunities to study collective quantum phenomena and exotic phase configurations and transitions. Recently  a variety of magnetic phases and frustrated configurations of spin models have been realised using directional control of couplings. For example, the XY model on a triangular lattice of atomic condensates has been studied with an independent coupling control along only two directions \citep{Struck2011}, realised via an elliptical shaking of the lattice. In addition, simulations of the Ising and XY models have been demonstrated in trapped ions with direction dependent couplings created by collective vibrational modes \citep{TrappedAtoms2004}. 

To realise a fixed dissipative structure consisting of dissipative channels and barriers, the following experimental approaches may be promising.
Since polaritons are quasi-particles that exist in semiconductor microcavity environments, their excitonic or photonic components can be directly accessed by manipulating quantum wells or microcavities, respectively. 
The technique of implanting protons into the quantum wells or into the top of distributed Bragg reflectors \citep{MFraser2} makes independent spatial control of both the exciton and the cavity photon energies possible, which in turn leads to a local control over the polariton decay rate. The spatial control of the polariton lifetime, i.e. the dissipation profile $\gamma({\bf r})$, can therefore be fabricated with the proton implant technique with a multi-layer mask \citep{MFraser2}.
Another way to access the exciton states alone is to create a controlled stress by applying a pin to the back side of the substrate \citep{Snoke2007}. As a result, a spatial trap is formed directly under the stressor where polaritons have an energy minimum and the cavity photon states and the exciton states are strongly coupled. Away from the center of the trap the lowest polariton states are almost purely photon-like which makes the coupling of the exciton states and cavity photon states weaker. This in turn means that the lifetime of polaritons at high energy is shorter than the lifetime of those at the energy minimum. Thus, in principle, such strain-induced traps can be used to create a configuration of dissipative channels and barriers if the tip radius of the pin can be decreased to a micrometer or less. 

The dynamic dissipation control for realising dissipative gates can be achieved by electrical carrier injection which leads to localized losses 
due to excited state absorption and bimolecular annihilation involving
polarons and long-lived triplets \cite{RoadMap}. Alternatively, the local control of the dissipation can be achieved by 
increasing the biexciton formation rate. Biexcitons can be created by two-photon absorption, by exciton absorption, or by inducing polariton-biexciton transitions \citep{biexcitons_1,biexcitons_2,biexcitons_3,biexcitons_4}.

To relate the characteristics of the dissipative control with the interaction between the condensates we first study a configuration of two polariton condensates. The $\pi$ phase modulation between geometrically fixed condensates is achieved by increasing the amplitude of the dissipative gate between condensates, as shown in Fig.~\ref{Figure2}(a). The complete transition to the $\pi$ phase state happens at higher gates in presence of higher dissipative barriers meaning that the excitonic landscape with deeper valleys supports stronger couplings between condensates. Here ``strong coupling" means that the condensates have to remain synchronised after the dissipative gate of a particular amplitude is placed. Such strong coupling can be achieved by, for example, using a uniform pumping profile or by closely arranging the condensates with large intersections of their individual pumpings. Next we consider the same close arrangement of pumpings for a $3 \times 3$ square block of polariton condensates with the pumping profile as in Fig.~\ref{Figure2}(b). In this case, the dissipative gates are placed between the vertical stripes of condensates (see Fig.~\ref{Figure2}(c)). The initial gateless state is configured to be ferromagnetic as shown in Fig.~\ref{Figure2}(d),  The amplitude of dissipative gates is then increased which leads to spin configurations in Fig.~\ref{Figure2}(e-g) with an antiferromagnetic coupling between vertical stripes of condensates in the final configuration. 
Here we note that the both chosen dissipative and pump profiles serve the same purpose of preventing undesired interactions. The former destroys polaritons by decreasing their lifetime, i.e. increasing losses, and the latter creates exciton reservoirs that block polariton outflows due to repulsive exciton-polariton interactions.

\begin{figure}[t!]
\centering
\includegraphics[width=8.6cm]{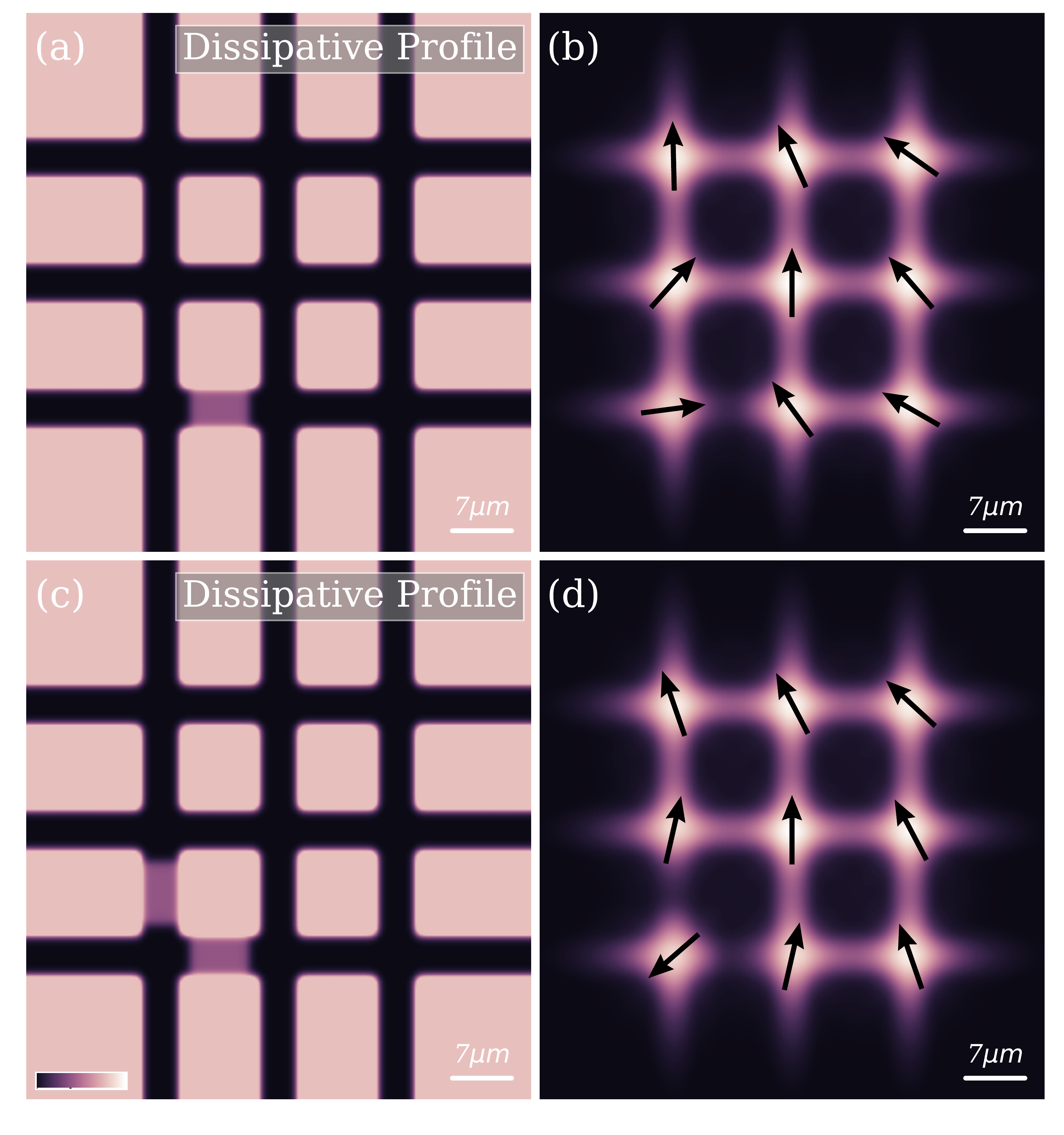}     	
\caption{(a) A dissipation profile structure similar to that of Fig.~\ref{Figure2}(b) but with only one dissipative gate. This dissipative gate makes the particular link antiferromagnetic while all other links remain ferromagnetic and, thus, frustration is created in the network. The resulting spin configuration is shown in (b). (c) The addition of another dissipative gate removes frustration from the system and leads to the spin configuration which is shown in (d).  The dissipative gate's amplitude is $\gamma_{\rm gate} = 7$, and other simulation parameters are the same as in Fig.~\ref{Figure2}.}
 \label{Figure3}
\end{figure}

\begin{figure}[t!]
\centering
\includegraphics[width=8.6cm]{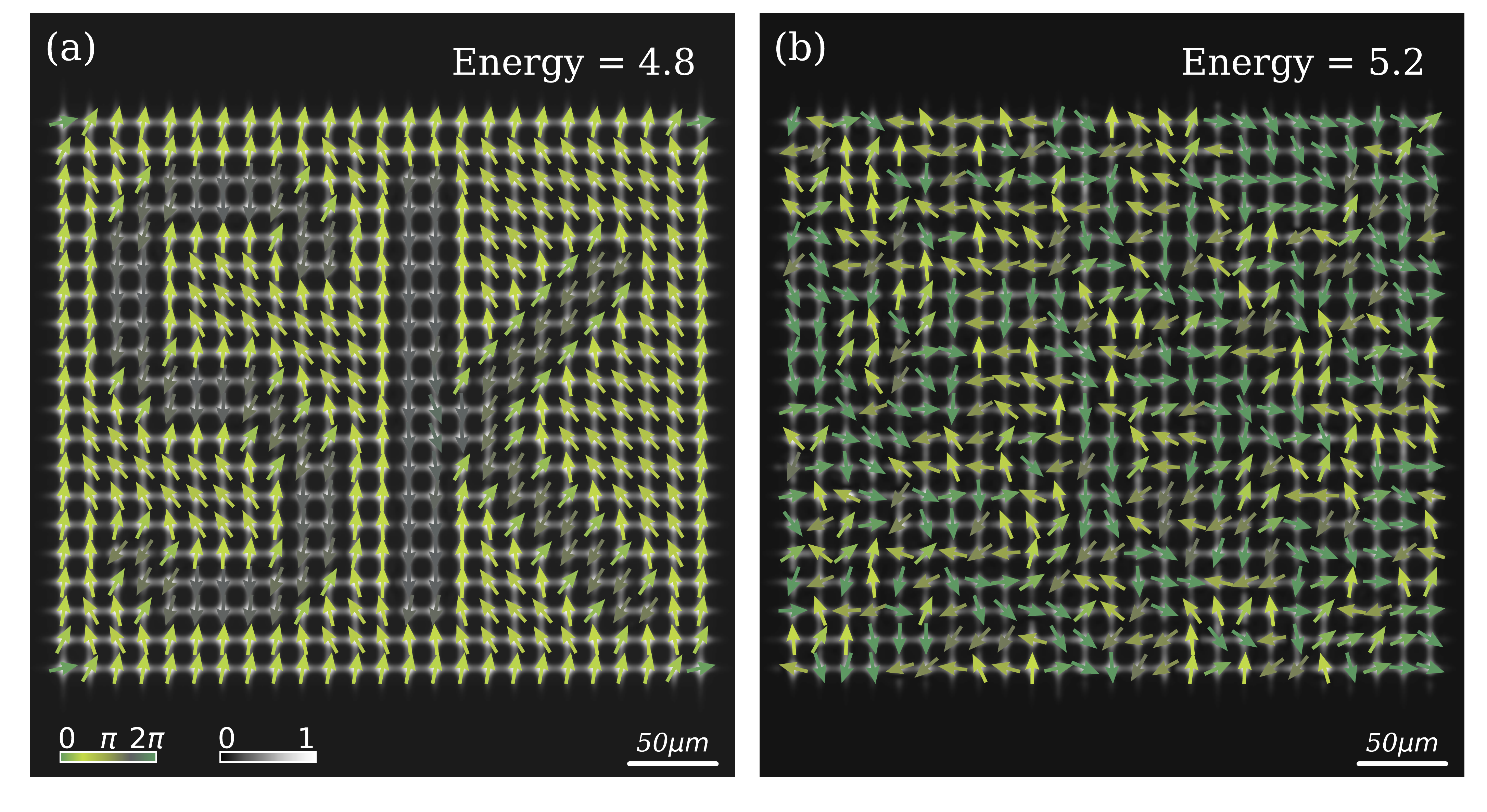}     	
\caption{The $20 \times 25$ square lattice of polariton condensates with the ground state shown in (a) and one of the excited states in (b). The polariton densities are shown with black-and-white colour scheme, the phase differences are plotted with a cyclic green-grey colour scheme. The dissipative gates $\gamma_{\rm gate} = 7$ are placed between the condensates constituting letters ``Sk" and other condensates to create antiferromagnetic couplings, while all other couplings are ferromagnetic. The ground state spin configuration (a) resembles ``Sk" letters while the excited state (b) is blurred. The energies are in dimensionless units. The dimensionless parameters for numerical simulations of (\ref{e1},\ref{e2}) are $b_0 = 0.1$, $P_0 = 60$, $\gamma_{\rm barrier} = 60$, $\gamma_{\rm gate} = 8$, $d = 3$, while other parameters are the same as in Fig.~\ref{Figure2}.}
 \label{Figure4}
\end{figure}

To check the stability of an arbitrary network, we next demonstrate that the individual control of couplings can be realised. Figure \ref{Figure3}(a) shows the dissipative profile with only one dissipative gate placed for the bottom-left condensate. This dissipative gate creates frustration in the network and makes this particular coupling antiferromagnetic while all the other couplings are ferromagnetic. The resulting spin configuration is depicted in Fig.~\ref{Figure3}(b) and demonstrates how the frustration spreads across the network. The addition of another dissipative gate as in Fig.~\ref{Figure3}(c) removes frustration from the system and leads to the spin configuration which is shown in Fig.~\ref{Figure3}(d). The absence of perfect alignment of the spins in this case, i.e. the deviations from $0$ and $\pi$ phase differences, we associate with the varying density of condensates in the lattice. We also note that perfect alignment can be expected for a network realising the $XY$ model while for polaritonic networks many other models including Kuramoto and Stuart-Landau can be realised under different conditions, as we discussed in  \citep{OurLattices2019}.

For a potential implementation of RC or analogue Hamiltonian optimisation it is important to demonstrate the scalability of the polaritonic network with dissipative gates, channels, and barriers. In Fig.~\ref{Figure4} we show a configuration of 500 condensates with 92 dissipative gates, which are placed so that regions with $\pi$ phase difference can form arbitrary symbols, in this case ``Sk". While the excited states do not maintain any recognisable spin configuration (see Fig.~\ref{Figure4}(b)), the lowest energy state indeed corresponds to the letters ``Sk" (see Fig.~\ref{Figure4}(a)).

In conclusion,  we  have shown how to engineer a lattice of nonequilibrium condensates with a pairwise control of classical spin interactions. The former is achieved by a spatial variation of a dissipation profile which is the key to control and direct the flows of polaritons. Such localised control of pairwise spin interactions paves the way to the study of previously intractable complex physical systems and to the manufacturing of new materials. It adds another dimension to the flexibility and tuneability of control parameters in lattice spin models. 
Polaritonic networks with spatially dependent dissipation profiles can be considered as promising candidates for reservoir computing, analogue Hamiltonian optimisation, logic devices, 
They give full access to the study of large-system phase transitions with arbitrary ferromagnetic or antiferromagnetic couplings, and symmetry-breaking caused by frustration. 
These networks can be used as laboratory realisations of new and exotic states of matter. In our paper, we focused on a polaritonic network, however, other nonequilibrium physical systems can also be used, including photon condensates as well as coupled lasers.

\section*{Acknowledgements}
N.G.B. acknowledges financial support from the NGP MIT-Skoltech. K.P.K. acknowledges the financial support from Cambridge Trust and EPSRC.
The authors are grateful to Jeremy Baumberg, David Snoke and St\'ephane K\'ena-Cohen for useful discussions, Samuel Alperin and Alexander Johnston for proofreading, and  to Sofia Berloff for making Fig.~\ref{Figure1}.

\end{document}